\documentstyle[aps,floats]{revtex}
\advance\voffset by 0.5in
\begin{document}
\draft
\newcommand{\bce}{\begin{center}} \newcommand{\ece}{\end{center}}
\newcommand{\beq}{\begin{equation}} \newcommand{\eeq}{\end{equation}}
\newcommand{\beqy}{\begin{eqnarray}}
\newcommand{\eeqy}{\end{eqnarray}} \input epsf
\renewcommand{\topfraction}{0.8}
\twocolumn[\hsize\textwidth\columnwidth\hsize\csname
@twocolumnfalse\endcsname


\title{Post-inflationary thermalization with hadronization scenario}
\author{Prashanth Jaikumar and Anupam Mazumdar} 
\address{Physics Department, McGill University, Montr\'eal, 3600,
Qu\'ebec, Canada H3A 2T8.}
\maketitle

\begin{abstract}

We study thermalization of the early Universe when the inflaton can
decay into the Standard Model (SM) quarks and gluons, using QCD
arguments. We describe the possible formation of the thermal 
plasma of soft gluons and quarks well before the completion of
reheating. Relevant interaction rates of leading order processes and
the corresponding thermalization time scale is presented. We discuss
hadronization while thermalizing the decay products of the inflaton,
when the reheat temperature of the Universe is below the QCD phase
transition but above the temperature of the Big Bang nucleosynthesis.
\end{abstract}

\vskip2pc]

\section{Introduction}

After the end of inflation, which is required to solve some of the
outstanding issues of hot Big Bang cosmology \cite{linde90}, the
homogeneous inflaton starts oscillating around its minimum. During
this period the average oscillations of the inflaton mimics the
equation of state of a pressureless fluid. The inflaton oscillations
persist until the inflaton decays and fragments in order to provide a
bath of relativistic species in kinetic and chemical equilibrium.  In
the literature the temperature of the relativistic thermal bath
acquired from the {\it complete} decay products of the inflaton is
known as the {\it reheat temperature}: $T_{\rm rh}$. The equilibration
of the bath takes finite time and the whole process is known as {\it
thermalization} of the inflaton decay products \cite{kolb90}.  The
time scale of thermalization depends sensitively on the inflaton
coupling to the decay products, which is usually not known when the
inflaton is considered to be a SM and $SU(3)$ gauge singlet,~\cite{olive90}.

Note that it is important for Big Bang nucleosynthesis (BBN) that at
the time when the electroweak interaction between neutrons and protons
freezes out (around the temperature of few MeV) the Universe must be
dominated by the SM relativistic degrees of freedom. Therefore it
becomes important that the inflaton must couple to the SM fields if
not directly but via non-renormalizable interactions \footnote{
Thermalization, reheating, and curvature perturbations may arise from
the known sector of the minimal supersymmetric SM (MSSM) flat
directions \cite{enqvist02c}, whose coupling to the SM quarks and 
leptons are known, see \cite{enqvist02}.}.

Although a thermal bath of temperature $\sim 1$~MeV is necessary, 
there is no direct evidence of a thermal history beyond the BBN era.
Therefore, the reheat temperature could in principle lie anywhere between 
$m_{\phi}\geq T_{\rm rh}\geq {\cal O}(1)$~MeV, where $m_{\phi}$ is the inflaton
mass. In supersymmetric (SUSY) theories there is an upper bound on reheat 
temperature arising from thermal production of the superpartner of the 
graviton: the gravitino~\cite{sarkar96,ellis84}. For the gravitino mass 
$\sim 100$~GeV, $T_{\rm rh}\leq 10^{9}$~GeV \cite{ellis84} (assuming 
that the Universe expands adiabatically from $T_{\rm rh}$ onwards).

The issue of thermalization has been addressed in many papers
\cite{ellis87,dodelson88,enqvist90,enqvist93,enqvist94,zimdahl97,mcdonald00,allahverdi00}
and very recently some concrete ideas have been put forward in
\cite{sarkar00,allahverdi02}. In earlier papers
\cite{ellis87,dodelson88,zimdahl97} elastic interactions, such as
$2\rightarrow 2$ scattering and annihilation have been considered,
while in \cite{enqvist90,enqvist93}, it was shown that elastic
collisions lead to kinetic equilibrium while redistributing the energy
density, and $2\rightarrow 3$ (particle number changing) processes
that lead to chemical equilibrium take a longer time. In
\cite{mcdonald00}, it was suggested that elastic scattering followed
by prompt decay might lead to rapid thermalization. The new idea
behind rapid thermalization was proposed in \cite{sarkar00}, where
inelastic scattering such as $2\rightarrow 3$ process have been
invoked from the very beginning.  It was found that thermalization
time scale is roughly given by the inverse of the inelastic scattering
rate $\Gamma_{\rm inel}^{-1}$. In this paper, we consider
thermalization with a hadronization scenario, where $T_{\rm rh}$ might
be below the QCD scale $\Lambda_{QCD}\sim 0.2$~GeV.  Especially when
the inflaton scale is as low as $m_{\phi}\sim {\cal O}(1)$
TeV~\cite{randall95,mazumdar99}, and the reheat temperature is below
the QCD scale $\sim 200$~MeV, then the importance of thermalization
and hadronization of the quark gluon plasma (QGP) becomes an
interesting issue.  Further note that the inflaton decay products
still have energy ranging from ${\cal O}(m_{\phi})$ down to $T_{\rm
rh}$. The hard hitting quarks and gluons must lose their initial
momentum towards the last stages of reheating.  Depending on whether
there is a hadronic bath, or a soft bath of quarks and gluons, the
inflaton decay products will lose their energy differently.  Our main
goal in this paper is a qualitative understanding of 
thermalization and hadronization process in a cosmological context
with an order of magnitude estimations.

\section{ inflaton decay to quarks and gluons}

After inflation has ceased, the inflaton field $\phi$ begins to
execute coherent oscillations about the minimum of its potential. 
The perturbative decay occurs over several such oscillations
\footnote{The initial stages of the inflaton oscillations could give
rise to non-thermal production of bosons and fermions, usually known
as preheating \cite{traschen90,dolgov90}. Alternatively the inflaton
condensate could fragment due to running mass
\cite{enqvist02a,enqvist02b}.  It is important to highlight that
preheating does not give rise to thermalization. Preheating produces
multi particle non-thermal spectrum at initial stages which still
requires to be thermalized with chemical and kinetic
equilibrium.}. The initial stage of reheating begins right after
$H_{inf}\simeq m_{\phi}$, where $H_{inf}$ denotes the Hubble expansion
of the Universe towards the end of inflation, and $m_{\phi}$ denotes
the mass of the inflaton. Note that the inflaton oscillations always
dominate the energy density until the inflaton decay is completed.
\beq H^2(a)\simeq\frac{\rho_{i}}{3 M_{\rm
P}^2}\left(\frac{a_{in}}{a}\right)^3\sim \Gamma_{\phi}^2\,, \eeq where
$\Gamma_{\phi}$ is the inflaton decay rate, and $\rho_{i}$ is the
energy density stored in the inflaton sector, and $a$ is the scale
factor of expansion. The subscript $in$ denotes the initial time which
can be taken to be the onset of inflaton oscillations. The inflaton
oscillations dominate until $\tau\sim \Gamma^{-1}_{\phi}$.  The
inflaton energy goes into a thermal bath of relativistic particles
whose energy density is determined by the reheat temperature $T_{\rm
rh}$, given by
\begin{equation}
\label{reheat}
T_{\rm rh} \sim g_{\ast}^{-1/4}\left(\Gamma_{\phi}M_{\rm P}\right)^{1/2}\,,
\end{equation}
where $g_{\ast}$ is the number of relativistic degrees of freedom. The above
estimation is simple and it still stands as the only valid estimate 
of the reheat temperature.

As $\phi$ is a gauge singlet with respect to all gauge symmetries
under consideration, it follows that for the case of QCD with quarks,
it can couple only to color singlets, and to SU(2) gauge singlets
constructed from the left-handed quark field which forms a doublet in
the standard model. In this paper we mainly concentrate upon 
interactions 
\beq {\cal L}_{\phi g g}=\frac{\phi}{M_{\rm P}}
F^{\mu\nu}F_{\mu\nu}
\eeq 
which describes the perturbative decay of the inflaton to gluons. Inflatons could also decay into SM quarks through
\beq {\cal L}_{\phi q q}=\frac{\phi}{M_{\rm P}}
(H\bar{q}_L)q_R\,, 
\eeq 
where $H$ is the Higgs doublet, $q_{L}$ is the SU(2) doublet, and 
$q_{R}$ is singlet. The neutral Higgs scalar eventually decays into the SM 
quarks and leptons via Yukawa couplings.

We can now estimate the number density of hard hitting quarks 
and gluons at time: $\tau>\tau_{in}$, where $\tau_{in}$ is the 
reference time when the inflaton starts oscillating. The number 
density is given by 
\beq
n_{\chi}(\tau)=2n_{\phi}(a_{in})(1-{\rm e}^{-\Gamma_{\phi}(\tau-\tau_{in})})
\biggl(\frac{a_{in}}{a}\biggr)^3\,,
\label{ninf}
\eeq
where $n_{\phi}(a_{in})=\rho_{\phi}(a_{in})/m_{\phi}$, and we collectively 
designate hard hitting quarks and gluons by $\chi$. Note that we are
neglecting here the particle number changing processes in the above 
estimation.


\section{Formation of the thermal bath of soft quarks and gluons}

The inflaton decay rapidly builds up a large number density of quarks
and gluons. Note that fragmentation or hadronization cannot occur in
the early stages of the inflaton decay, the formation of bound states
being prevented by the IR cutoff imposed by the finite expansion rate
of the universe. The time scale of expansion during inflaton oscillations is 
$t_{\rm exp}\sim 1/H(\tau)\geq 1/m_{\phi}$ which is much smaller than the 
typical time scale of strong interactions $\tau_s\sim$ GeV$^{-1}$
(for $m_{\phi}\gg {\cal O}(1)$~GeV). However, these 
mechanisms will be important towards the end of the inflaton decay, 
and we will discuss these in section V.

We now ask the question as to how this dense system of hard quarks and
gluons will thermalize, what the corresponding temperature would be
and estimate the time to reach the final reheat temperature $T_{\rm
rh}$. A rigorous study of thermalization would involve solving
numerically the relativistic Boltzmann equation for the time evolution
of the quark and gluon densities, with splitting functions
incorporated from pQCD, and consequent logarithmic corrections to the
quark and gluon densities. A full calculation of the parton cascade
evolution needs to be done in the expanding space-time background,
which is beyond the scope of the present work. It is worthwhile at 
this stage to review and recast the arguments for rapid thermalization 
that appear in references \cite{sarkar00,allahverdi02} in a QCD 
based approach.

The mechanism therein involves creating a soft thermal bath of
gauge bosons from scattering of hard particles accompanied by the
emission of a soft gluon from one of the legs of the $2\rightarrow 2$ 
diagram. This is the gluon Bremsstrahlung. The emitted soft gluon can now
be involved in a scattering process with another quark (or gluon),
thereby leading to an exponential growth in the number density of soft
particles. Due to pair annihilation processes, the thermal bath in 
QCD will be composed of soft quarks as well as soft gluons. The soft 
quarks and gluons can form a thermal plasma with an instantaneous 
temperature \cite{kolb90} 
\beq
\label{inst}
T_{inst}\sim \biggl(g_*^{-1/2}H(\tau_{inst})\Gamma_{\phi} M_{\rm P}^2
\biggr)^{1/4}\,,
\eeq  
where $H(\tau_{inst})\geq \Gamma_{\phi}$ is the Hubble parameter at 
the time when the soft thermal bath is created. This instantaneous 
temperature reaches its maximum $T_{max}\leq m_{\phi}$ soon after 
the inflaton field starts to oscillate around the minimum of its potential. 
Also an important point to note here is that since the inflaton energy 
density is still dominating 
$\rho_{\phi}\approx H^2M_{\rm P}^2\sim a^{-3/2}$, the 
instantaneous temperature  falls as 
\begin{equation}
T_{inst}\sim a^{-3/8}\left(g_{\ast}^{-1/2}\frac{\Gamma_{\phi}}{H}\right)^{1/4}
\,,
\end{equation}
instead of $T_{inst}\sim a^{-1}$. Numerical simulations also support 
this argument\cite{chung99}. In the next section we estimate 
$T_{max}$.

For elastic processes, one cannot reduce the average energy of the
system quickly since no new particles are created (we are ignoring
the red-shift of the particle momenta for the moment). The
Bremsstrahlung process redistributes the energy among softer
constituents and these soft particles act as a very efficient sink of
energy once the bath starts to form. An analogous mechanism for
heavy-ion collisions was suggested by Shuryak~\cite{shuryak92} that a
thermally equilibrated quark-gluon plasma (QGP) would be created in a
two-step process (``hot-glue scenario''), wherein the gluonic bath
would be created in a much shorter time than it took the quarks to
thermalize.

We now proceed to make some quantitative estimates for the
cross-sections and scattering rates.  Note that these are order of
magnitude estimates, and more accurate methods such as using Boltzmann
equations with collision terms taken from perturbative QCD may be
developed as has been done for the finite size system in heavy-ion
collisions~\cite{wong96}. However, in 
the cosmological context, it is adequate to first obtain an order of
magnitude estimate for a proposed physical scenario.


The $2\rightarrow 3$ processes and higher order ($2\rightarrow n$)
gluon branching will lead to the formation of a plasma of soft quarks
and gluons. When this is eventually formed, exchanged gluons will be
screened over a distance scale $\mu^{-1}\sim (1/gT)$ corresponding to
the momentum scale $gT$~\footnote{The magneto static gluons are not
screened perturbatively, and this scale does not apply to them. If
magnetic screening effect is also taken into account then
$\mu^{-1}\sim (1/g^2T)$. However, for our estimation purposes, a
starting point is to take the minimal form for the screening of the
gluon propagator.}  \cite{lebellac}. In the case of quarks being
exchanged in a plasma, we can use the cutoff given by the kinetic
theory which gives the quarks an effective mass of ${\cal O}(gT)$. A
more consistent perturbative approach in a thermal plasma would be to
the Braaten-Pisarski re-summation scheme~\cite{braaten90}, which
incorporates effective propagators and vertices for the quark and
gluons, and amounts to a reordering of perturbation theory. Since we
are interested only in order of magnitude estimations here, we choose
the bare propagators and minimal expressions for the cutoff.

For a sufficiently dilute system of quarks and gluons, the concept of
a thermal plasma does not make sense, in which case, we choose to
regulate in the infra-red by using, at any instant, the inverse of the
inter-particle separation of the decay products of the inflaton (hard
quarks and gluons which we denote by $\chi$). This is determined by
Eq.~(\ref{ninf}) at any instant of time. For $\Gamma_{\phi}\leq H(\tau)$, 
we expand the right hand side of the expression Eq.~(\ref{ninf}) in terms 
of $\Gamma_{\phi}/H$, and with the help of Eq.~(\ref{inst}), we immediately 
obtain the number density $n_{\chi}$ of hard quarks and gluons, and from it, 
the infra-red cutoff
\begin{equation}
n_{\chi}^{-1/3}\sim \left(\frac{T_{inst}^4}{m_{\phi}}\right)^{-1/3}\,.
\end{equation}
It should be noted that the expressions for energy transfer will be
less divergent than those for the absolute cross-section since the
energy transfer also tends to zero for very soft scattering. This
reduces the efficacy of soft processes for thermalization.

As emerged in studies of thermal and chemical equilibration in 
heavy-ion collisions, inelastic processes ($2\rightarrow n$, where $n>2$) 
are likely to be the dominant mechanism for rapid thermalization and 
entropy generation. Within perturbative QCD itself, it has been shown 
that processes that are higher order in $\alpha_s$ can be more important 
for equilibration. Thus, gluon branching from a quark or gluon line must 
be taken into account.

To achieve thermalization at a temperature well below $m_{\phi}$, one
needs to generate many soft particles so that while the energy density
remains constant, the number density rapidly increases. Then the
average energy per particle also decreases. The creation of a soft
gluonic bath can proceed via Bremsstrahlung emission of a gluon from a
quark or gluon participating in a scattering process ($2\rightarrow 3$
process). The energy loss due to Bremsstrahlung can be mitigated by
the Landau-Pomeranchuk-Migdal (LPM) effect~\cite{lpmigdal} in a dense
medium. This effect has been studied in a QED~\cite{moore} 
as well as in a QCD plasma, with significant modifications in the energy 
loss profile of a parton jet traversing a QCD plasma~\cite{wang93}. If 
the mean free path of the incident parton is small compared to the 
formation time of the emitted gluon (which can become large for small 
angle scattering by uncertainty principle arguments), interference 
effects between multiple scattering has to be taken into account. 
This issue has been addressed in calculations of photon and gluon 
production rates from an equilibrated QGP, as well as in the phenomenon 
of jet quenching. The importance of this effect can be gauged as follows. 

Suppose the incident quark of energy $E_q$ emits at a small angle
$\theta$, a gluon ($k^{\mu}=(\omega_g,k_z,k_{\perp})$) with energy
$\omega_g\sim T_{inst}$. 
The formation time of the emitted gluon would be given by the uncertainty 
principle as~\cite{wang93} 
\beq
\tau(k)\sim
\frac{2\omega_g}{k_{\perp}^2}\sim \frac{2}{\omega_g\theta^2}\,, 
\eeq
where $\theta\approx k_{\perp}/\omega_g$. The ratio of the mean free
path to the formation time is given by 
\beq
\frac{L}{\tau}\sim\frac{1}{n_{\chi}\sigma_{\rm inel}}
\frac{\omega_g\theta^2}{2}\,,
\label{lpm}
\eeq 
where $\sigma_{\rm inel}=\sigma_{\chi\chi\rightarrow \chi\chi g}$ is 
the inelastic cross-section. It has a logarithmic dependence on $t$ 
from the Bremsstrahlung emission, and a $1/t$ dependence from the exchanged 
momentum. The extra emitted gluon gives one extra power of $\alpha_s$ 
compared to the elastic case leading to   
\beq
\sigma_{\rm inel}\sim\frac{\alpha_s^3}{t_{\rm min}}{\rm log}
\biggl(\frac{m_{\phi}^2}{t_{\rm min}}\biggr)\,.  
\eeq 
Substituting the above equation into Eq.~(\ref{lpm}), and using the fact that
$t_{\rm min}\sim g^2 T_{inst}^2$ and $n_{\chi}\sim T_{inst}^4/m_{\phi}$, 
we find 
\beq
\frac{L}{\tau}\sim\frac{m_{\phi}}{g^2~T_{inst}~{\rm log}
\biggl(\frac{m_{\phi}^2}{g^2 T_{inst}^2}\biggr)}\,.
\label{landau}
\eeq 
If this ratio is much less than one, the LPM effect causes
suppression of energy loss, and consequently, the hard quarks (or
gluons) lose their energy at a slower rate. This can weaken the
functional dependence of $dE/d\tau$ on the incident energy~\cite{wang93}, 
leading to a suppression of energy loss, and an increase in thermalization 
time. From Eq.~(\ref{landau}), it is evident that for a given 
$4\pi\alpha_s\ll 1$, the assumption of additive energy loss from successive 
scattering applies, since the fraction $T_{inst}/m_{\phi}$ is much less 
than unity implying that $L/\tau\gg 1$ (the logarithm cannot counteract 
the dependence from the prefactor). 

Continuing with the inelastic process in the limit $L/\tau\gg 1$
(additive energy loss), the time scale for an inflaton decay product
to lose an energy $\sim m_{\phi}$ by such processes turns out to be
much smaller than in the elastic case, as was shown
in~\cite{sarkar00}. The reason is that the soft gluon population
(neglecting its annihilation to fermions) grows exponentially since
each soft gluon produced radiatively also acts as a subsequent
scattering center for the production of further soft gluons. The bath
will be actually composed of a certain number of fermions as well, due
to the forward and backward annihilation processes between quarks and
gluons. However, for our estimation purposes, it only matters that
hard quarks/gluons are scattering off soft particles, whose total
number is important though not its species content. Due to the
(almost) exponential increase in the number of soft particles, we
estimate that the rate of energy loss is given by \beq
\frac{dE}{d\tau} \approx n_g\sigma_{\rm inel}T_{\rm inst}\,, \eeq
where $n_{g}$ is the number density of the soft quarks (and gluons)
$n_{g}\sim g_{\ast}T_{inst}^3$~\footnote{We are assuming, as
in~\cite{sarkar00}, that the average energy loss is $\sim
T_{inst}$. This is not strictly true if one considers the collinear
behavior of the Bremsstrahlung process, but we can ignore
 log($m_{\phi}/T$) corrections which only speed up the thermalization rate
by a numerical factor of order one.}

The inelastic scattering rate is then given by \beq \Gamma_{\rm
inel}=\frac{1}{E}\frac{dE}{d\tau}\sim
\frac{g_{\ast}\alpha_s^2T_{inst}^2}{4\pi~m_{\phi}}~{\rm log}\biggl(
\frac{m_{\phi}^2}{g^2T_{inst}^2}\biggr)\,.  \eeq Note that if we were
to consider higher order gluon emission (2 or more emitted gluons),
$\Gamma_{\rm inel}$ would be suppressed by further powers of
$\alpha_s$ and will not compete with the $2\rightarrow 3$
process. Furthermore, emission of gluons softer than $T_{inst}$, which
would ultimately require handling collinear divergences are not
considered since those ultra soft gluons would have to be re-scattered
back up to a momentum of $T_{inst}$. As we are interested only in the
time scale to create a thermalized plasma at roughly $T_{inst}$, this
aspect will not concern us here~\cite{sarkar00}. We will return to the
point about the $2\rightarrow n$ processes at the end of this
section. Continuing with our estimate, the time to lose an energy
$m_{\phi}$ is then \beq \tau_{\rm inel}=\Gamma_{\rm inel}^{-1}\sim
\frac{m_{\phi}}{\frac{dE}{d\tau}}
\sim\frac{4\pi~m_{\phi}}{g_{\ast}\alpha_s^2~T_{inst}^2~{\rm log}
\biggl(\frac{m_{\phi}^2}{g^2 T_{inst}^2}\biggr)}\,.\label{tinel}
\eeq 

Note that this result is a factor $4\pi\alpha_{s}/g_{\ast}$ different than
previous estimates \cite{sarkar00}, because the lower cutoff is taken
as the screening mass rather than the typical temperature (which would
be the energy of the individual constituents themselves, and would be
ignorant of many-body effects). 
Now we are able to estimate the largest temperature of the
instantaneous thermal bath before reheating. By using 
$H(t_{inst})\simeq \Gamma_{\rm inel}$, Eq.~(\ref{inst}), and 
$\Gamma_{\phi}=g_{\ast}^{1/4}T_{\rm rh}^2/M_{\rm P}$, we obtain the maximum
instantaneous temperature 
\begin{equation}
\frac{T_{max}}{T_{\rm rh}} \sim \left(g_{\ast}^{3/4}\alpha_{s}^2\frac{M_{\rm P}}
{4\pi~m_{\phi}}\right)^{1/2}\,.
\end{equation}
Note that $T_{max}$ is couple of magnitudes larger than the reheat 
temperature, but still smaller compared to the inflaton mass: 
$T_{max}\leq m_{\phi}$.

It is sufficient for us to estimate the thermalization time by 
studying $2\rightarrow 3$ processes, since further gluon branching 
($2\rightarrow n, n>3$) will come with additional factors of 
$\alpha_s\ll 1$. The lower scattering cross-section for such 
processes renders them higher-order corrections to the inelastic 
scattering rate and the thermalization time scale. This conclusion 
can change if the mean free path of the hard particles becomes 
comparable to the formation time of the emitted gluon, i.e., if 
the LPM suppression is severe.


\section{Towards hadronization}

Thus far, we have described a generic scenario towards reheating the
Universe without bothering too much on the reheat temperature itself. An
implicit assumption was made that the reheat temperature was sufficiently 
above the QCD scale. However, a priori there is no reason that the 
Universe could not be reheated at a temperature lower than the QCD scale and
above the BBN temperature. Such a scenario is plausible if the inflaton 
scale is sufficiently low. Keeping the reheat temperature 
$T_{\rm rh}< \Lambda_{QCD}\approx 200$~MeV, we will concentrate upon the
hadronization issue.

Naturally if the thermal bath of soft quarks and gluons is formed near 
$T_{inst}\sim T_c$, where $T_c$ is the critical temperature 
($\sim 200$ MeV), the quarks and gluons will rapidly hadronize. 
However, if the thermal bath is initially formed at a much higher 
temperature than a GeV, it will hadronize at a later time when
the expansion of the universe has red-shifted the energy down to the
confinement scale. This is particularly
important if thermalization occurs within a Hubble time $H^{-1}$.  

Let us imagine that there already exists a plasma of soft quarks and gluons.
Since the temperature of the plasma is inversely related to the scale factor 
of the FRW metric as $T\propto a^{-3/8}$. It follows that during the inflaton
oscillation dominated phase 
\beqy 
\frac{T(\tau)}{T(\tau_0)}&=&\biggl(\frac{\tau_0}{\tau}\biggr)^{1/4}\,,
\quad\nonumber \\
\frac{s(\tau)}{s(\tau_0)}&=&\frac{T^3(\tau)a^3(\tau)}{T^3(\tau_{0})
a^3(\tau_{0})}=\left(\frac{\tau}{\tau_{0}}\right)^{5/4}\,.
\label{entropy}
\eeqy 
where $\tau_{0}$ denotes any reference time.
It is interesting to mention here that similar relations are derived 
for the evolution of the relativistic plasma of quarks and gluons 
formed in heavy-ion collisions~\footnote{There, it is appropriate 
to adopt Bjorken's idealized hydrodynamical scenario of the quark-gluon 
plasma as an expanding relativistic fluid~\cite{bjorken83}.
Approximate longitudinal boost invariance and the ideal equation of 
state for a relativistic fluid imply that the temperature and entropy 
density depend on the proper time $\tau$ as 
${T(\tau)}/{T(\tau_0)}=({\tau_0}/{\tau})^{1/3}$, and
${s(\tau)}/{s(\tau_0)}=({\tau_0}/{\tau})$.}.

At time $\tau=\tau_c$ when $T=T_c\sim 200$~MeV which is the
critical temperature for hadronization, the fluid is dilute enough
that strong non-perturbative forces recombine the quark-gluon soup
into colorless hadrons. This constitutes a phase transition whose
nature depends on the number of light quark flavors (from universality
arguments~\cite{wilczek84}) and the quark masses (which act as
external fields). Assuming the strange quark mass to be light on the 
$\Lambda_{QCD}$ scale, we can say that the hadronization proceeds 
by a first order phase transition with bubbles of hadronic matter 
appearing within the quark-gluon plasma, which grows in number till
the end of the phase transition~\cite{shuryakbook}. The time for this 
transition to occur is denoted by $\tau_h$; the hadronization time.

It is possible to estimate $\tau_{h}$ as follows~\cite{CYKbook}. 
Throughout the phase transition, the temperature remains constant 
at $T_c$, while entropy is decreased in going from the quark-gluon 
to the hadronic phase. This means that Eq.~(\ref{entropy}) cannot be
applicable in describing the phase transition. It is helpful to imagine
the quark gluon plasma as a subsystem where the temperature remains
$T(\tau_0)=T(\tau)=T_{c}\sim 200$~MeV for the duration of the phase 
transition. Then, the ratio of the entropy densities is simply the ratio of
$\epsilon + p$ of the plasma, which scales as the fourth power of temperature. 
From Eq.~(\ref{entropy}), it follows that 
\begin{eqnarray}
\label{entropy1}
\frac{s(\tau)}{s(\tau_c)}&=&\frac{\epsilon(\tau)+p(\tau)}{T_{c}}\cdot
\frac{T_{c}}{\epsilon(\tau_{c})+p(\tau_{c})}\, \nonumber \\ &=&
\left(\frac{\tau_c}{\tau}\right)\,,\quad
\tau>\tau_c \,,
\end{eqnarray} 
In the case of heavy-ion collisions, a rerun of the above steps yields 
the relation ${s(\tau)}/{s(\tau_c)}=({\tau_c}/{\tau})^{4/3}$.

This means that the phase transition in the early universe occurs
at a slower rate than in heavy-ion collisions, since the
entropy/particle is some fixed quantity in the two different
phases. We will see that this fact is borne out by our final
expression for the hadronization time. During the transition, the
system can be described by a mixed phase, with a fraction $f(\tau)$ in
the quark-gluon phase ($1-f(\tau)$ in the hadronic phase). Since
entropy is additive, we find using Eq.~(\ref{entropy1}), 
\beqy
f(\tau)&=&\frac{1}{s_{qg}(T_c)-s_h(T_c)}\biggl(\{f(\tau_c)s_{qg}(T_c)+
[1-f(\tau_c)]\nonumber \\
&&s_h(T_c)\}\times\frac{\tau_c}{\tau}-s_h(T_c)\biggr) 
\eeqy
The entropy densities in the two phases can be traded for the respective 
degeneracies ($g_{qg}\sim 37$ and $g_h\sim 3$) to give 
\beq
f(\tau)=\frac{1}{g_{qg}-g_h}\left({f(\tau_c)g_{qg}+[1-f(\tau_c)]g_h}
\frac{\tau_c}{\tau}-g_h\right)\,.
\eeq 
At $\tau=\tau_h$, the quark-gluon phase is completely hadronized. Thus, 
it follows from the above equation that 
\beq
\tau_h=\left[\frac{g_{qg}}{g_h}f(\tau_c)+1-f(\tau_c)\right]\tau_c 
\label{hadtime}\,.
\eeq 
Starting with quark-gluon matter at $f(\tau_c)=1$, we find that
$\tau_h\sim 10\tau_c$. In heavy-ion collisions, this relation gives
$\tau_h\sim 6\tau_c$~\footnote{In the heavy-ion case, there is a power 
of $3/4$ for the term within square brackets in Eq.~(\ref{hadtime}), 
which follows from the different scaling law of entropy density with time. 
The fractional power implies a reduced hadronization time in heavy ion 
collisions.}.

Coming back to Eq.~(\ref{entropy1}), we observe that $\tau_c$ denotes the 
time for the plasma to cool down to the critical temperature for 
hadronization $T_c$. The origin of time is taken to be the moment 
when the inflaton begins its oscillations. Taking 
$\tau_{\rm inf}\sim m_{\phi}^{-1}$, we can estimate an upper bound 
on $\tau_c$ as 
\beq 
\tau_c\leq\Gamma_{\phi}^{-1}\sim m_{\phi}^{-1}\alpha_{\phi}^{-1}\,,
\eeq 
if the universe is to be reheated below the QCD phase transition. Here, $\alpha_{\phi}$ is the non-renormalizable coupling of the inflaton
to the SM quarks and leptons. For a low reheat temperature, a small inflaton coupling as set by the above equation is therefore necessary. It 
is known that this time in the lab frame must be on the $\mu$s time scale.

Unlike for the case of heavy-ion collisions, where hadrons, once formed, 
decouple chemically and thermally from each other after a certain time 
and stream freely to the particle detectors, that may not be the case 
here. Since the hadronization time is short compared to the inflaton 
decay lifetime, it follows that the (hard) inflaton decay products will 
continue impinging on the newly-formed hadrons and cause break-up due 
to the large energy transfer. How efficient this process is depends 
on the number density of these hard particles. One also needs some 
estimate of the rate of energy lost by these hard particles 
traveling now in a hadronic phase rather than a soft gluonic bath. 
It is interesting to compare this mode of energy loss for the hard 
particles to the previously considered inelastic scattering process. 
We also expect fragmentation to be important for lowering the average energy, since energetic partons can now fragment into a large number of less energetic hadrons, which can re-scatter inelastically among themselves, or resonance decay into lighter hadrons. We comment on the relative importance of these processes
below.

\section{Final stages of hadronization}
\subsection{Deep-inelastic scattering (DIS)}

The break-up of hadrons by hard-hitting quarks can be assessed through
DIS. The structure functions at some $x,Q^2$ can be obtained from the
parton distributions that appear in the DGLAP equation~\cite{GLAPD} \beq
\frac{df_i(x,Q^2)}{d~{\rm
log}~Q^2}=\frac{\alpha_s}{2\pi}\int_x^1\frac{dy}{y}P_{ij}(z)f_j(y,Q^2)
\eeq where the momentum fraction of the daughter parton $x=zy$, with
$y$ the momentum fraction of the parent quark. $P_{ij}(z)$ are
splitting kernels that could be written down in lowest order pQCD, for
example. Rather than perform a detailed DGLAP evolution, we wish to
estimate the time scale for the hard quarks to lose the bulk of their
energy. From Eq.~(\ref{ninf}), we can conclude that the number density
of hard particles at temperature of order the hadronization scale
($\sim 200$ MeV) is very low ($n_q\leq T^4/m_{\phi}\sim 10^{-6}$
fm$^{-3}$), for an inflaton mass of $m_{\phi}=10^5$ GeV (relatively
small inflaton mass which gives rise to reheat temperature smaller
than $\Lambda_{QCD}$) and the temperature being around the QCD phase
transition. At the same temperature, the number of hadrons is $n_h\sim
(m_hT_h)^{3/2}{\rm e}^{-m_h/T}$.  The Boltzmann suppression is least
for the lightest hadron, namely pions, whose mass $m_{\pi}=140$ MeV is
very close to $T_h$, where the hadronization temperature $T_h\simeq
150$ MeV. Therefore $n_h\sim T_h^3e^{-1}\sim 10^{-1}$ fm$^{-3}$. This
implies that while not many hadrons break up, the hard quarks can lose
energy quite efficiently due to the large number of participants. The
average fractional energy loss of a hard quark per hard collision is
not much less than 1/2 or 1/3. The mean free path of the hard quark is
approximately $l\sim(n_h~\sigma_{DIS})^{-1}$. Assuming a typical
cross-section of micro-barns for the integrated deep-inelastic
cross-section at a momentum transfer $Q\sim 1$
GeV~\cite{Muta}~\footnote{Although, the possibility of larger momentum
transfers also implies a much shorter time scale, since the exchange
of high energy gluons (at $Q^2\gg 1$ GeV $^2$) would include an
enhancement factor of $\alpha_s^2/\alpha_{\rm
em}^2$~\cite{halzen84}.}, we find $l\sim 10^4$ fm. Since these quarks
are relativistic, this implies that the hard quarks would lose most of
their energy within a time $\sim l/c\approx10^{-19}$s.

\subsection{Fragmentation and hadron-hadron scattering}

The mechanism of energy loss by 2$\rightarrow$ 3 inelastic scattering
can still occur, though it is not correct to use one gluon exchange as
the only interaction when quarks are so dilute. If we continue to use
the perturbative estimate, at the temperature $T_h$, from
Eq.(~\ref{tinel}), we obtain $\tau_{\rm inel}\sim 10^{-11}$s. However,
we believe that this would not be an accurate estimate because of the
strong non-perturbative forces active at this separation scale of the
quarks. In light of these strong non-perturbative forces, we may
consider the fragmentation process whereby the energy of the quark is
degraded by conversion to hadrons. Late thermalization and
hadronization can happen also by fragmentation of the hard quarks followed by
hadron-hadron scattering.  For purposes of estimation, we neglect the
effect of the hadronic medium and using a typical proper time of
formation $\langle\tau\rangle\simeq 1-2$ fm/c, the time in the lab
frame (time dilated) is given by~\cite{Feyn78} \beq t_{\rm
frag}\sim\langle\tau\rangle\frac{E_q}{m_h} \eeq Since the energy of
the hard quark $E_q\sim m_{\phi}\sim 10^5$~GeV and the hadronic mass
is about $0.1$~GeV, we find $t_{\rm frag}\sim 10^{-18}$s.  Compared to
the time scale for the DIS process, this is only slightly slower, and
the two processes might be comparable within a more detailed
treatment. The main difference between them is the dependence of the
time scale on the inflaton mass (logarithmic for DIS, and linear for
fragmentation). For $m_{\phi}\sim 10^5$ GeV, they are almost the
same. Since the late hadronization happens in a time that is very
short compared to the inflaton decay lifetime, complete hadronization
is achieved only when all inflatons have decayed. It should be noted
however that fragmentation alone is not sufficient since one must also
take into account that one produces hadrons with a large multiplicity
and with energies considerably above $T_h\sim 0.1$~GeV. These hadrons
will inelastically scatter among themselves, and lose energy. Since
these integrated cross-sections ($\sigma_{hh}$) are of order
millibarns, using the formula $l=1/(n_h\sigma_{hh})$, it is found that
the time to reach the equilibrium temperature of the hadron soup is
of order $10^{-22}$s. This is much smaller than the time for the
initial fragmentation, so the overall conclusion is that one must
compare the fragmentation process against the deep inelastic
process. For an inflaton mass $m_{\phi}\sim 10^5$ GeV, this leads to
the picture that the late inflaton decay products thermalize via an 
initial parton shower, followed by DIS, then
fragmentation followed by hadron-hadron scattering, within a time much shorter
than the inflaton decay lifetime.


\section{Conclusion}

We have studied the perturbative decay of the
inflaton to quarks and gluons, focusing particularly on the aspect of
thermalization and energy loss of these hard particles in an expanding
universe. For a given inflaton mass $m_{\phi}$, the coupling has to be
small enough in order that thermalization is achieved before inflaton
decay is completed. The thermalization time is short, driven
principally by 2$\rightarrow$ 3 inelastic processes. After the thermal
plasma is formed, it continues to cool in the background of an
expanding universe, until a critical temperature for hadronization is
reached. The laws of expansion are different than those of a
relativistic fluid expanding on account of it's own pressure, as is
the case with heavy-ion collisions.\vskip 0.1cm

While we have not examined in detail the 
hadronization process, which occurs in a mixed phase at roughly 
a constant temperature and is accompanied by a reduction in entropy, 
we have estimated that the hadronization time is likely to be on the 
order of micro-seconds as well ($\sim 10\mu$s) based on the relationship between the maximum temperature of the 
plasma and the reheat temperature, which should be below 
$\Lambda_{QCD}$. 

This state of hadrons may not be immediately stable, however, 
on account of the small cool-down and thermalization time in comparison
to the inflaton decay lifetime. Hard inflaton decay products will 
impinge on this bath of hadrons, and cause some break up into quarks and 
gluons once again. One can also safely conclude, on account of the small 
hadronization time, that complete formation of a hadronic bath 
will be stable only when all inflatons have decayed and the number 
density of hard particles is negligible. The temperature characterizing 
that state of the thermal hadronic bath composed of non-relativistic 
particles can be considered to be the reheat temperature $T_{\rm rh}$. 
                         
In this paper, we have borrowed ideas and expressions from
relativistic heavy-ion physics to discuss cosmological issues such as
the thermalization of the inflaton decay products. This intermingling
is natural especially near the hadronization scale where current
experiments are active, thus the issue of low reheat temperature for a
QCD plasma could be successively improved by a more rigorous treatment
of the early non-equilibrium stages prior to thermalization.

\section*{Acknowledgment}

The authors are thankful to Charles Gale for discussions. P. J
acknowledges support from the Natural Sciences and Engineering
Research Council of Canada. A. M is a Cita-National fellow.


\end{document}